\documentclass[runningheads]{cl2emult}

\usepackage{makeidx}  
\usepackage{graphicx} 
\usepackage{subeqnar} 
\usepackage{multicol} 
\usepackage{cropmark} 
\usepackage{eso}      
\makeindex            

\begin{document}
\title*{The connection between X--ray Clusters and Star Formation}
\toctitle{The connection between X--ray Clusters
\protect\newline  and Star Formation}
%
%
\titlerunning{X--ray clusters and star formation}
%
\author{Paolo Tozzi\inst{}
\and Colin Norman\inst{}}
\authorrunning{Tozzi \& Norman}
%
%
\institute{The Johns Hopkins University, 3400 N. Charles, 21218, Baltimore, MD}

\maketitle              

\begin{abstract}

The properties of X-ray clusters of galaxies can be well understood in
terms of a competition between shock heating and adiabatic
compression.  Strong shocks are expected to be important for massive
clusters, while adiabatic compression is dominant for small clusters
and groups.  The scale of the shock/adiabatic transition is marked by
a change of slope of the $L$--$T$ relation and in the global properties of
the emitting plasma.  This scale is connected to star formation
processes.  Two quantities are crucial: the {\sl average energy}
injected in the IGM from stars and SNe, and the {\sl epoch} of the energy
injection. We show how these quantities can be synthesized in terms of
specific entropy, which ultimately determines the X-ray emission from
groups and clusters.

\end{abstract}

\section{The properties of X--ray clusters of galaxies}
Clusters of galaxies are objects of great cosmological relevance.  In
fact, they are believed to constitute a well--defined population of
X--ray sources with simple evolutionary behaviour (see, e.g.,
\cite{betal} and references therein).  However, their X--ray emission
properties are not fully understood: the simple self--similar scaling
predicts a relation between the X--ray luminosity and the temperature,
$L\propto T^2$ \cite{k86}, which is at variance with the observed
relation $L\propto T^3$.  There is further steepening of the $L$--$T$
at temperatures around $1$ keV \cite{pon}.  Moreover, a sharp
transition in the global properties of the emitting plasma is evident
below $1$ keV \cite{r97}. This behaviour has been related to the
presence of an entropy minimum in the intergalactic medium (IGM)
which, in turn, can be adiabatically compressed in the cluster
potential \cite{eh}\cite{k91}\cite{ba} or accreted through shocks, as
suggested by numerical simulations and analytical models \cite{cmt}.
Here we present a simplified argument to show how shock and adiabatic
compression compete in a manner regulated by the entropy level of the
IGM.  The entropy production (before accretion) is due to
non--gravitational heating by stars and SNe, and is determined by the
average energy and the epoch when such energy is released \cite{pcn}.

\section{The L-T relation in terms of entropy}

Here we present a simple argument to understand the $L$--$T$ relation
in terms of entropy production. The mimimum entropy of the diffuse IGM
is due to non--gravitational energy injection $kT_*$ at the epoch
$z_*$, when the gas density was $\rho_*\propto (1+z_*)^3$.  The
specific entropy initially is then $S\propto {\rm log} (K_*)$, where
$K_*\equiv kT_*/\mu m_H \rho_*^{2/3}$.  If, in the following, the IGM
is only adiabatically compressed, the final relation between density
and temperature is $\rho\propto (T/K_*)^{3/2}$.  The X--ray emissivity
due to bremsstrahlung is $\epsilon_{ff}\propto \rho^2T^{1/2}$.  The
central total luminosity of the cluster can be written:
\begin{equation}
L\propto \epsilon_{ff} V\propto T^{7/2}K_*^{-3} M\propto T^5K_*^{-3}\, ,
\label{eq1}
\end{equation}
where $V$ is the emitting volume.  Then, in the limit of pure
adiabatic compression, the $L$--$T$ relation is naturally expected to
be steeper than the self similar case (see also \cite{ba}).

On the other hand, for a generic IGM distribution we can write $K$ as:
\begin{equation}
K=T/\rho^{2/3}=(T/T_*) (\rho/\rho_*)^{-2/3}K_*\, .
\end{equation}
If we neglect further adiabatic compression, the quantity $K$ is
determined as follows.  In the case in which all the gas is accreted
through a strong shock, the ratio $(\rho/\rho_*)$ is simply the maximum
compression factor allowed, $(\rho/\rho_*)\rightarrow 4$. If the gas
is isothermal, we have $K\propto K_* T$.  Substituting the
actual value of $K$ for $K_*$ in equation \ref{eq1}, we obtain
$L\propto T^2$.  Thus, in the limit of strong shocks the self similar
scaling is recovered.

Of course it is too simplistic to describe the observed $L$--$T$
relation in terms of two power--laws.  In the real world, the
transition between the adiabatic and the shock regime is gradual.
However, the scale of the onset of the shocks is an important
quantity, and, from observations, is expected to occur around $1$ keV.
In the following section we show how this scale depends on the
entropy, which requires two pieces of information: {\sl how much} energy
is supplied, and {\sl when}.

\section{The synergy of X--ray and optical observations}

The mass scale at which shocks start to appear, is the scale for which
the infalling velocity of the accreting baryons is larger than the
sound speed $v_s=(\gamma kT/m_H\mu)^{1/2}$, where $\gamma = 5/3$ and
$\mu=0.59$ for a primordial IGM.

If the potential well is deep enough, the infall velocity is
essentially the free fall velocity $v_{ff}$ (i.e., the baryons follows
the dark matter).  However part of the work done by the gravitational
field goes into compression of the gas.  As a result, the baryons are
delayed with respect to the dark matter and the infall velocity $v_i$
of the gas is smaller than $v_{ff}$.  If we approximate the adiabatic
external flow with a Bernoulli equation \cite{tn}, the condition for
the formation of the shock is $v_i>v_s$. This occurs when the
gravitational term (or in other words, the free fall velocity) is
dominant with respect to the pressure term.  At the virial radius,
where the gas is accreted and reaches the maximum velocity, this
condition can be well approximated by:
\begin{equation}
{{v_{ff}^2}\over{c_s^2}} > \Big( {{\gamma +1}\over {\gamma -1}}\Big)
-{2\over{\gamma
-1}}\Big({{\rho_{ta}}\over{\rho_s}}\Big)^{\gamma -1}
\label{cond}
\end{equation}
where $\rho_s$ and $\rho_{ta}$ are the densities of the gas at the
shock and at the turn--around.  Assuming the shock occurs at
the virial radius and using mass conservation, we have $\rho_s = f_B
\rho_{dm} v_{ff} /v_i$.  Expanding
in $\rho_{ta}/\rho_{s}$, at the first order we have $v_{ff}\geq (1.87
- 1.95)c_s$, when $\rho_{ta}/\rho_s\simeq 1/8$ and $1/30$.

\begin{figure}
\includegraphics[width=6cm, height =4.2cm]{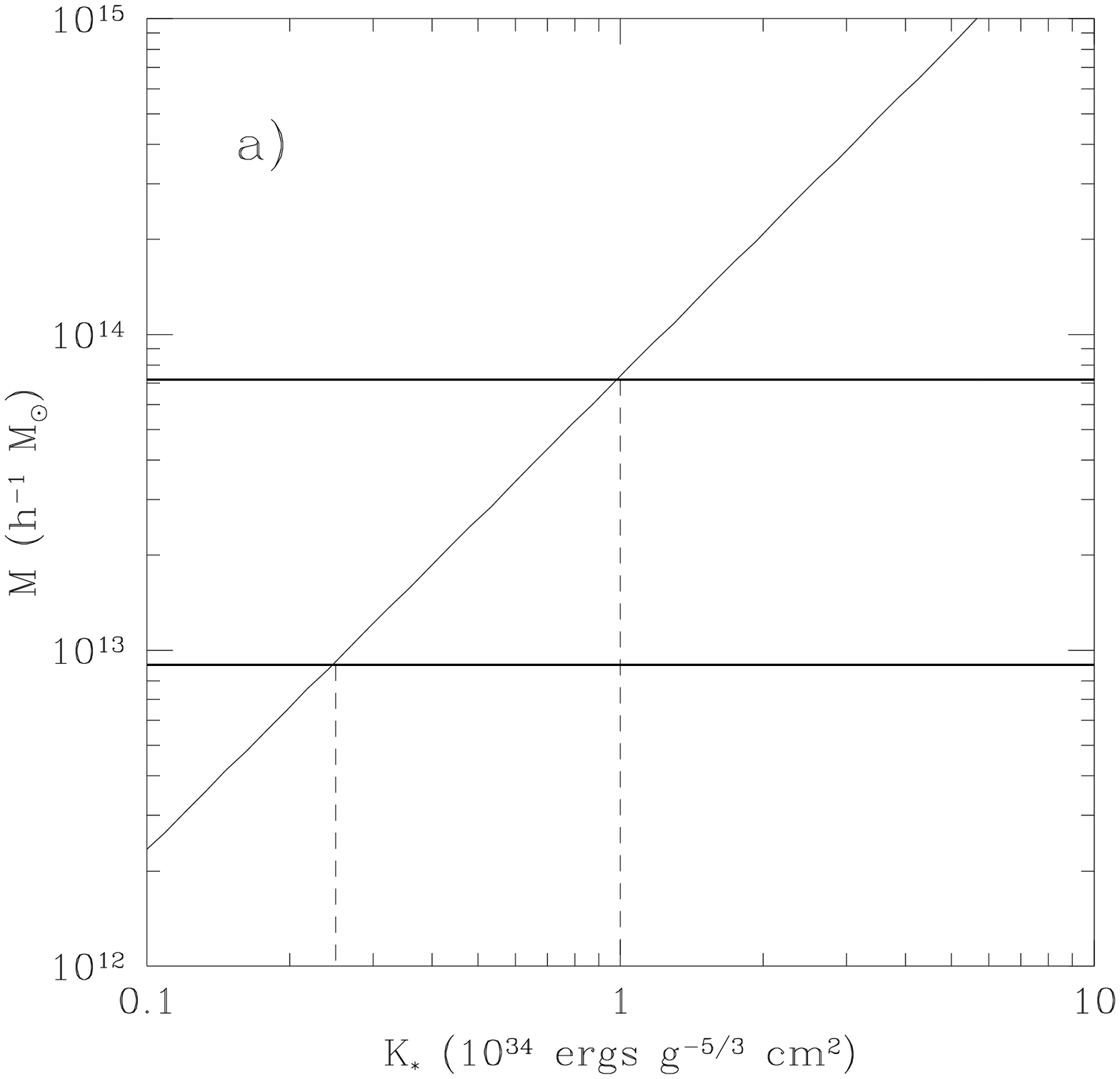}
\includegraphics[width=6cm, height =4.2cm]{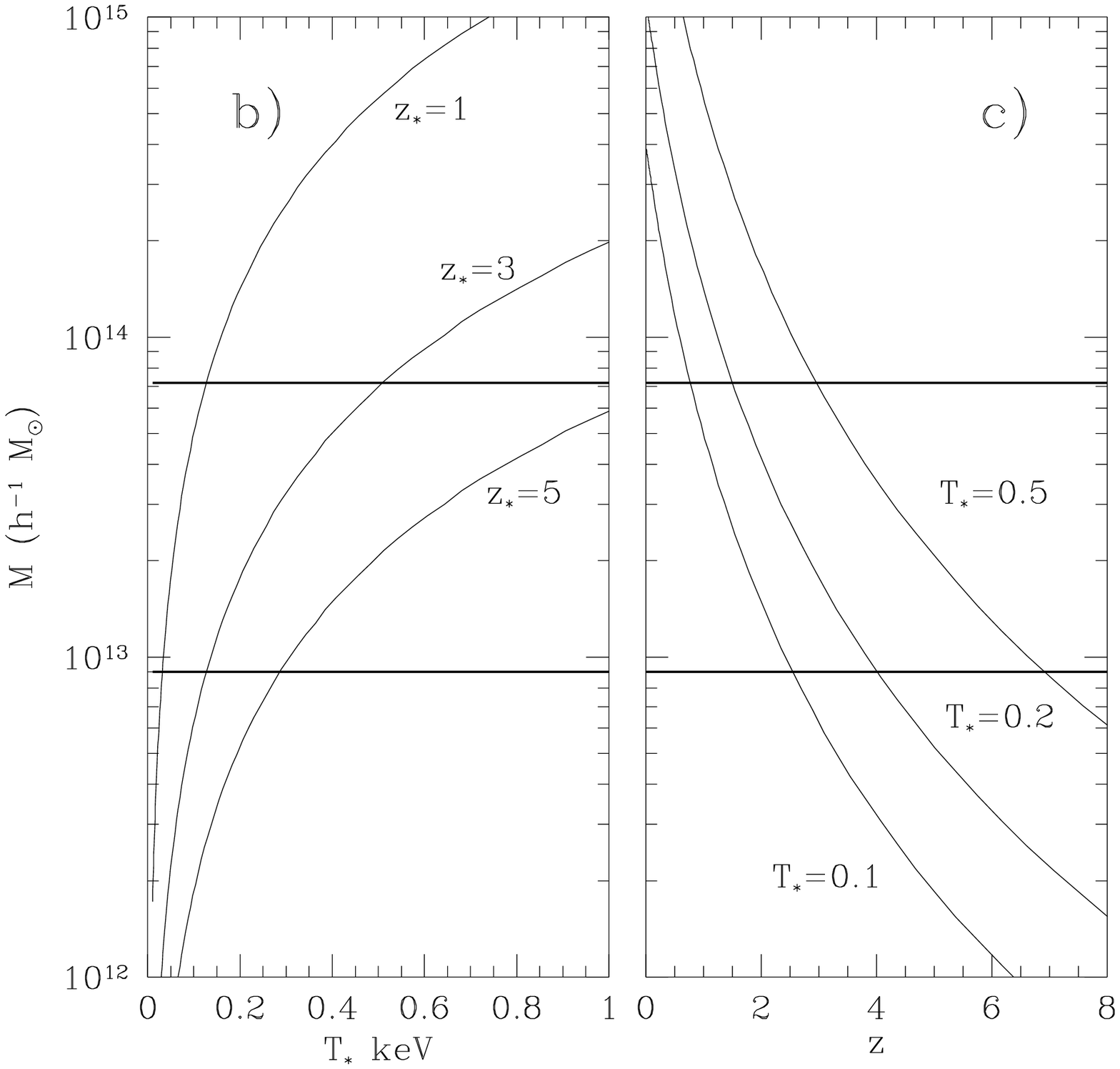}
\caption{Panel a): the mass scale of the adiabatic/shock transition as a
function of $K_*$; b): the same as a function of
the non gravitational energy scale $T_*$ for three different
redshifts $z_*$; c): the same as a function of redshift for three different
$T_*$ }
\label{fig1}
\end{figure}

If we assume that the entropy minimum $\propto {\rm log}(K_*)$ is
produced at a single epoch when the IGM was uniform (before the
formation of clusters and groups), we have $K_* = kT_* /\mu m_H
\rho_o^{2/3} (1+z_*)^{2}$.  Since $c_s^2=\gamma
K_*\rho_s^{2/3}$, the mass scale of the shock/adiabatic
transition is determined by $K_*$, as shown in the panel a) of figure
\ref{fig1}.  The two horizontal lines bracket the mass range roughly
corresponding to $0.5-2$ keV, with an associated set of values for
$K_*$ in the range $0.2-1\, \times 10^{34}$ ergs
g$^{-5/3}$cm$^2$. Here we use an average $M$--$T$ relation and take
into account the fact that a fraction $\leq 1/2$ of the mass is
responsible for more than $80$ \% of the X--ray emission in groups.
This argument is analogous to the one presented in \cite{pcn}, based on the 
entropy core in clusters.  

In panel b) and c) we show the behaviour in terms of the two variables
$T_*$ and $z_*$ independently.  In panel b) we see that the
transition scale occurs in the same mass interval for a range of
initial temperature $kT_*=(0.1-0.5)$ keV.  If the same energy is injected
later the entropy and the predicted transition
mass scale are higher.  This is clear in panel c), where the
dependence on $z_*$ for a given energy input is shown. 
These results hold under the assumption that the gas was uniform at
the epoch of entropy production.  If, instead, the gas was at a
contrast $\delta_*$ with respect to the background density, the
relation between energy and epoch is $T_*\propto
(1+z_*)^2\delta_*^{2/3}$ for a given $K_*$.  This indicates why,
after the collapse, a larger energy input is needed to obtain the same
effect (i.e., the same $K_*$).

Our discussion shows that the entropy is the key parameter, and demonstrates
how both energy and time are equally important in shaping the X--ray
emission from large scale structure.  Such considerations are
mandatory to calculate the energy budget that is necessary to
reproduce the observed $L$--$T$ relation.

\subsection{What is still missing?}

The description of the X-ray properties in terms of entropy can be
much more detailed.  In particular, the complete density and temperature
profiles, including shock heating and adiabatic compression, can be
computed in different regimes and cosmological frameworks \cite{tn}.
However, to define statistically the population of local and high z
clusters of galaxies, we still need some important pieces of
information.

The injection of $T_*$ is a continuous process, eventually peaking at
redshift $z\simeq 2-3$.  The emitting properties of the ICM will
result from the competition between the timescales for dynamical
evolution and entropy production.  Since there is a large scatter in
the formation history of massive halos in all the hierarchical
cosmologies, the properties fluctuate with the epoch of observation
and the mass scale.  Such intrinsic scatter is one of the most
relevant characteristics of the observed $L$--$T$ relation.  The
accretion of already virialized, small halos is expected to occur with
the free fall velocity, since the bound, clumped gas is not slowed by
pressure effect.  Consequently, part of the ICM will be always
shocked, in an amount depending on the presence of small scale
structure.  Cooling flows are not included in the present treatment;
on the other hand, they are expected to be important, not only for
mass deposition in the center, but even for their contribution to the
X-ray properties.

In summary, the physics of the X--ray
emitting plasma in clusters of galaxies is well understood.  The most
meaningful parameter is the entropy, which is directly connected to
the energy and the temporal scales set by star formation processes.  These,
in turn, can be directly observed in the optical and infrared bands
from ground and space based telescopes.  In the near future, the
synergy between the X--ray facilities and optical telescopes like the
VLT, is expected to give powerful insights in the formation and
evolution of cosmic structures.

\clearpage
\addcontentsline{toc}{section}{Index}
\flushbottom
\printindex

\end{document}